\documentclass[12pt]{iopart}

\usepackage[utf8]{inputenc}

\usepackage{iopams} 
\usepackage{setstack} 

\usepackage{graphicx}
\usepackage{hyperref}

\newcommand{\BB}{\begin{eqnarray}}
\newcommand{\EE}{\end{eqnarray}}

\newcommand{\shalf}{\frac{1}{2}}

\newcommand{\NN}{\frac{1}{N}}

\newcommand{\HH}{{\cal H}}

\newcommand{\non}{\nonumber}

\newcommand{\la}{\lambda}
\newcommand{\La}{\Lambda}
\newcommand{\w}{\omega}
\newcommand{\p}{\rho}

\newcommand{\Sig}{\Sigma}
\newcommand{\ga}{\gamma}
\newcommand{\Ga}{\Gamma}
\newcommand{\Th}{\theta}

\newcommand{\ww}{\w+i\delta}

\newcommand{\bSig}{\overline{\Sig}}

\newcommand{\bD}{\widetilde{D}}

\newcommand{\krest}{\dagger}
\newcommand{\akrest}{a^\krest}
\newcommand{\bkrest}{b^\krest}
\newcommand{\akr}{\akrest}
\newcommand{\bkr}{\bkrest}
\newcommand{\kk}{{\bf k}}
\newcommand{\pp}{{\bf p}}
\newcommand{\kkk}{{{\kk}_0}}
\newcommand{\kz}{{k_0}}

\newcommand{\RR}{{\bf R}}

\newcommand{\mkk}{{\bf -k}}

\begin{document} 
\title[Antiferromagnet with two coupled antiferromagnetic sublattices in a magnetic field]{Antiferromagnet with two coupled antiferromagnetic sublattices in a magnetic field} 
\author{A V Sizanov$^{1,a}$ and A V Syromyatnikov$^{1,2,b}$}

\address{$^1$Petersburg Nuclear Physics Institute, Gatchina, St.\ Petersburg 188300, Russia} \address{$^2$Department of Physics, St.\ Petersburg State University, 198504 St.\ Petersburg, Russia}

\eads{\mailto{$^a$alexey.sizanov@gmail.com}, \mailto{$^b$syromyat@thd.pnpi.spb.ru}}

\date{\today}
\begin{abstract}
	
We discuss the magnon spectrum of an antiferromagnet (AF) in a magnetic field $h$ consisting of two interpenetrating AF sublattices coupled by the exchange interaction at $T=0$. One-ion easy-plane anisotropy is also taken into account. We calculate using the $1/S$ expansion the gap in the spectrum which is a manifestation of the order-by-disorder effect in this system and the optical magnon mode splitting. Both of these phenomena originate from the inter-sublattice interaction. We calculate also the gap value at $h\approx h_c$ in the leading order of the small parameter $(h_c-h)/h_c$ using the magnon Bose-Einstein condensation treatment, where $h_c$ is the saturation field. By comparing results obtained within these two approaches we conclude that the $1/S$ expansion gives a qualitatively correct result at $h\sim h_c$ even at large one-ion anisotropy but it overestimates the gap value. The application is discussed of these results to the actively studied AF of the considered type $\rm NiCl_2$-$\rm 4SC(NH_2)_2$ (DTN).

\end{abstract}

\maketitle

\section{Introduction} 

\begin{figure}
	\label{dtnpicture} \centering 
	\includegraphics[scale=0.35]{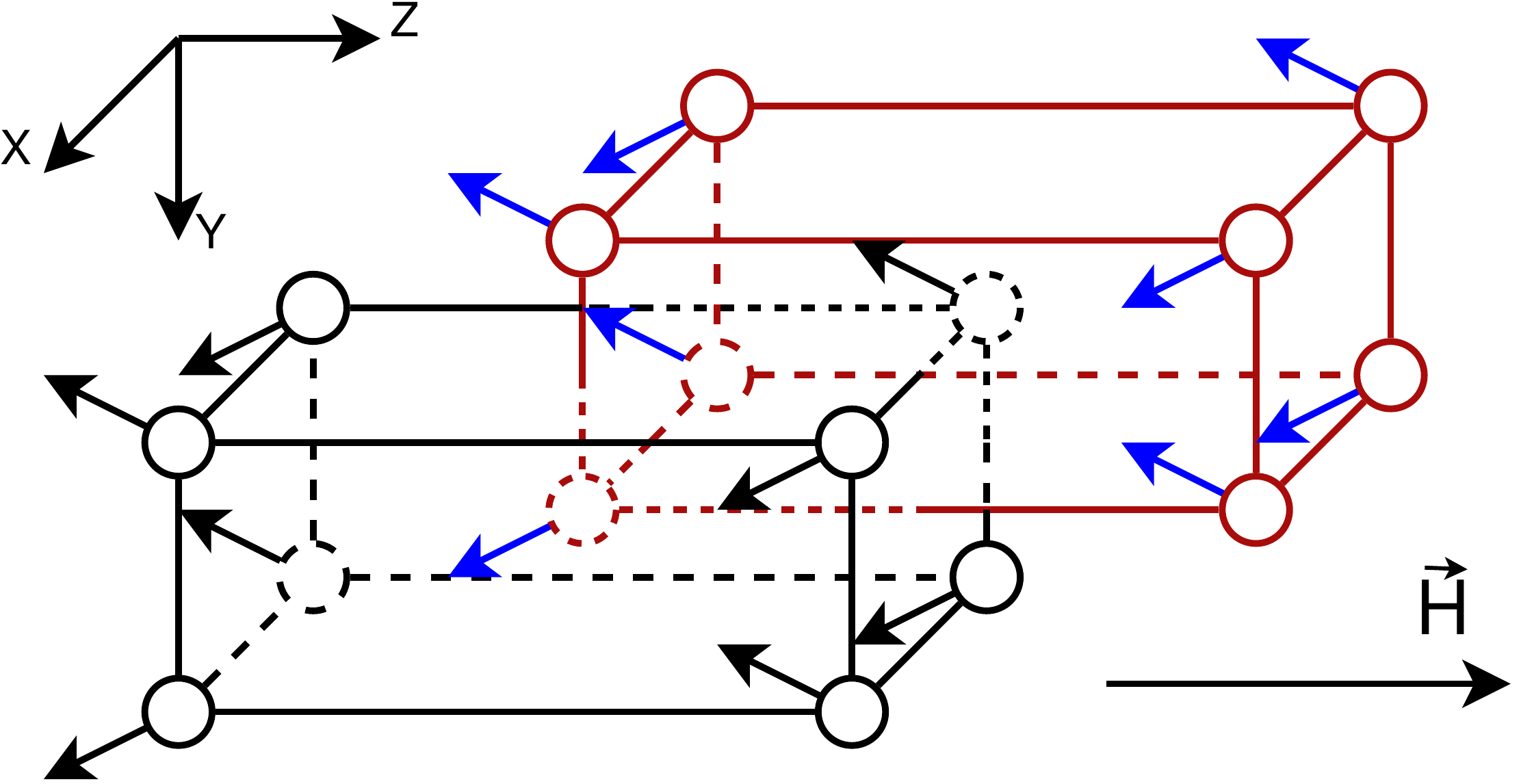} \caption{\label{affig} (Color online.) An antiferromagnet in magnetic field is presented containing two interpenetrating antiferromagnetic sublattices shown in different colors.} 
\end{figure}

Quantum criticality of magnetic systems driven by easily controllable external parameters like, for example, magnetic field and pressure, and properties of frustrated magnets have attracted much attention in recent years.	One of the most remarkable properties of frustrated systems with a continuously degenerate ground state is that fluctuations in them can make favorable some particular states via the so-called ‘order-by-disorder’ mechanism \cite{vil1,vil2,henley,shender}. An antiferromagnet (AF) containing two interpenetrating AF sublattices shown in \fref{affig} is an example of a system of this kind if atoms from one sublattice are located in a zero molecular field of atoms from another sublattice. It is the case, for example, if sublattices are coupled by exchange interaction. It was shown first by Shender \cite{shender} that quantum fluctuations lead to anisotropic corrections to the energy and collinear orientation of AF sublattices is selected in this way via the ‘order-by-disorder’ mechanism. This anisotropy is accompanied by a gap in one of the two initially degenerate Goldstone magnon modes. This gap was observed experimentally in the garnet $\rm Fe_2Ca_3(GeO_4)_3$ \cite{garnet}. 

Quantum critical properties of another magnet of this kind, $\rm NiCl_2$-$\rm 4SC(NH_2)_2$ known as DTN, have attracted much experimental and theoretical attention recently \cite{1,2,3,4,5,6,7,8,9,add1,add2}. DTN is characterized by the $I4$ space group with a body-centered tetragonal lattice that can be represented as two interpenetrating tetragonal AF sublattices and whose Hamiltonian has the form
\begin{eqnarray}
\label{ham} 
{\cal H} &=& {\cal H}_{SL_1}+{\cal H}_{SL_2}+{\cal H}_{int}, \\
&&\non\\
\label{hams}
{\cal H}_{SL_l} &=& \frac12 \sum_{i,j} J_{ij} {\bf S}_{i,l}{\bf S}_{j,l} +D\sum_{i} (S_{i,l}^z)^2+ h\sum_{i} S_{i,l}^z,
\end{eqnarray}
where ${\cal H}_{SL_l}$ is the Hamiltonian of the $l$-th AF sublattice, $l=1,2$ and ${\bf S}_{i,l}$ is the spin operator at site $i$ in $l$-th sublattice. The inter-sublattice interaction $\HH_{int}$ is considered to be small and is ignored in the majority of works devoted to DTN. Parameters leading to satisfactory agreement between theoretical calculations using Hamiltonian \eref{ham}, \eref{hams} and the majority of experimental data are $S=1$, $D=8.9$~K, $J_z=2.2$~K and $J_{x,y}=0.18$~K \cite{6}, where subscripts $x,y,z$ denote exchange constants along corresponding axes shown in \fref{dtnpicture}. Due to the strong one-ion easy-plane anisotropy the ground state at $h=0$ is disordered and all spins are predominantly in the $S^z=0$ state. That is why there are two quantum critical points in the magnetic field applied parallel to the hard $z$-axis. The first one which is located at $h=\tilde h_c = 2.1$~T separates a disordered and a canted antiferromagnetic phases. The second quantum critical point is at $h=h_c = 12.6$ T, where $h_c$ is the saturation field. It is between the canted AF and a collinear ferromagnetic phases.

A recent ESR experiment at $0<h<0.8h_c$ demonstrates the need to take into account a small inter-sublattice interaction in DTN \cite{1}. The model \eref{ham}, \eref{hams} with $\HH_{int}=0$ has a doubly degenerate gapless spectrum with one acoustic and one optical mode. It was found experimentally in \cite{1} that one of the two degenerate Goldstone modes has a gap and the optical mode is slightly split. To illustrate the idea that the inter-sublattice interaction is needed to describe the full set of experimental data, the authors of \cite{1} introduced a Dzyaloshinsky--Moria (DM) interaction between sublattices and described the gap qualitatively by a mean-field theory. The authors pointed out that an exchange interaction between sublattices should also be taken into account but they ignored it because it complicates calculations considerably and does not lead to the gap on the mean-field level. Thus, it looks desirable to reconsider the spectrum renormalization in the AF canted phase by taking into account the inter-sublattice exchange coupling because it also leads to the gap via the ‘order-by-disorder’ mechanism, as is described above, and the exchange coupling is normally much larger than the relativistic DM interaction \cite{moriya}.

This reconsideration is one of the present paper’s aims. We discuss below spectrum renormalization of the model \eref{ham}, \eref{hams} at $T=0$ in the first order of $1/S$ expansion and by using the magnon Bose-Einstein condensation (BEC) theory (at $h\approx h_c$) with the following inter-sublattice interaction:
\begin{equation}
	\label{hint} 
	{\cal H}_{int}=\sum_{i,j}  V_{ij} {\bf S}_{i,1}{\bf S}_{j,2}
\end{equation}
which is considered to be small, $V\ll J+D$. Using the $1/S$ expansion we obtain the optical mode splitting and expression for the gap which is a generalization of the well-known formula by Shender derived for $h=0$ \cite{shender}. The  optical mode splitting is of zeroth order in $1/S$ whereas the gap is of first order in $1/S$, being the result of quantum fluctuations. These results are inapplicable at small $h$ in the case of large $D$, when the ground state does not have the Neel order. The gap value is derived also at $h\approx h_c$ in the leading order of the small parameter $(h_c-h)/h_c$ using the magnon BEC treatment. We demonstrate by comparing results obtained within these two approaches that the $1/S$ expansion reproduces qualitatively the field dependence of the gap at $h\sim h_c$ even at large one-ion anisotropy $D$ but it overestimates the gap value. Particular estimations using the results of the $1/S$ expansion show that $V\sim0.1$ K is needed to describe the optical mode splitting and the gap obtained in DTN in recent ESR experiment \cite{1} at $h\approx0.8h_c$.

This paper is structured as follows. We analyze in \sref{secsw} the Hamiltonian \eref{ham}--\eref{hint} using the $1/S$ expansion. The neighborhood of the point $h=h_c$ is discussed using the magnon BEC theory in \sref{qcp}. Application to DTN of the results obtained in \sref{secsw} and \sref{qcp} is discussed in \sref{dtn}. \Sref{conc} contains our conclusions. Two appendices are added with some details of calculations.

\section{Spin-wave analysis} 
\label{secsw}

We assume in this subsection that the exchange coupling of AF sublattices is small: $V\ll D+J$. In this case spins in AF sublattices order parallel to each other in zero field \cite{shender} and they cant opposite to the field direction by an angle $\theta$ at finite $h$ smaller than its saturation value $h_c$ as it is shown in \fref{affig}. It is convenient to introduce a local coordinate frame $(x',y',z')$ in each lattice site in which the mean spin value is parallel to $z'$-axis. Spins components in the laboratory coordinate system $(x,y,z)$ are expressed as follows via those in the local coordinate frames:
\begin{eqnarray}
	\nonumber S^x_{n,1} &=& S^{x'}_{n,1},\nonumber\\
	S^y_{n,1} &=& S^{y'}_{n,1} \cos\Th+S^{z'}_{n,1} \exp(i\kkk\RR_{n,1})\sin\Th,\nonumber\\
	S^z_{n,1} &=& -S^{y'}_{n,1} \exp(i\kkk\RR_{n,1})\sin\Th+S^{z'}_{n,1} \cos\Th,\nonumber\\
	\label{reframe} S^x_{n,2} &=& S^{x'}_{n,2},\\
	S^y_{n,2} &=& S^{y'}_{n,2} \cos\Th+S^{z'}_{n,2} \exp\left[ i\kkk(\RR_{n,2}-{\bf Q}) \right] \sin\Th, \nonumber\\
	S^z_{n,2} &=& -S^{y'}_{n,2} \exp\left[i\kkk(\RR_{n,2}-{\bf Q}) \right] \sin\Th+S^{z'}_{n,2} \cos\Th, \nonumber 
\end{eqnarray}
where ${\bf Q}=(1/2,1/2,1/2)$ is a vector connecting a spin of one AF sublattice to the neighboring one of another AF sublattice, ${\bf k}_0=(\pi,\pi,\pi)$ is AF vector of an AF sublattice and we put the distance between neighbor spins in an AF sublattice to be equal to unity. Imaginary exponents in \eref{reframe} describe the Neel ordering in the XY-plane. We use the Holstein-Primakoff representation of spins components in the local coordinate frame written as 
\begin{eqnarray}
	S^{x'}_{n,1}+iS^{y'}_{n,1} &=& \sqrt{2S}\akr_n \sqrt{1- \akr_n a_n / 2S}\approx \sqrt{2S} \akr_n \left(1 - \akr_n a_n / 4S\right),\nonumber\\
	S^{x'}_{n,1}-iS^{y'}_{n,1} &=& \sqrt{2S}\sqrt{1- \akr_n a_n / 2S }\,\cdotp a_n\approx \sqrt{2S} \left(1- \akr_n a_n / 4S \right) a_n,\non\\
	S^{z'}_{n,1} &=& -S+\akr_n a_n,\non\\
	\label{HP} 
	S^{x'}_{n,2}+iS^{y'}_{n,2} &\approx& \sqrt{2S} \bkr_n \left(1- \bkr_n b_n / 4S\right),\\
	S^{x'}_{n,2}-iS^{y'}_{n,2} &\approx& \sqrt{2S} \left(1- \bkr_n b_n / 4S \right) b_n, \nonumber\\
	S^{z'}_{n,2} &=& -S+\bkr_n b_n,\non 
\end{eqnarray}
where operators $a$ and $b$ are introduced for the first and the second AF sublattices, respectively. Substituting \eref{reframe} and \eref{HP} into the Hamiltonian \eref{ham}--\eref{hint} and producing the Fourier transformation we obtain 
\begin{equation}
	\label{hamseq} {\cal H} = \sum_{i=0}^6{\cal H}_i, 
\end{equation}
where 
\begin{eqnarray}
	\fl 
	\frac1N{\cal H}_0 = S^2\left((2J_0+2\bD+V_0){\cos}^2\Th-2\frac{h}{S}\cos\Th-2J_0\label{h0}\right),\\
	\fl 
	\frac{1}{\sqrt N}{\cal H}_1 = i(a_\kkk-\akr_\kkk)\sqrt{\frac S2}\left[ S(2J_0+2\bD+V_0)\cos\Th-h\right] \sin\Th,\label{h1}\\
	\fl 
{\cal H}_2 = \sum_\kk \akr_\kk a_\kk\left[g_\kk{\cos}^2\Th +S(J_0+\bD)(1-3{\cos}^2\Th)- SV_0{\cos}^2\Th+h\cos\Th\right] \non\\
+ \frac S2 \sum_\kk (a_\kk a_\mkk+\akr_\kk \akr_\mkk) (J_\pp-\bD){\sin}^2\Th \non\\
+ \label{h2} \frac S2 \sum_\kk (\akr_\kk b_\kk+\bkr_{-\kk} a_{-\kk}) \left[V_\kk(1+{\cos}^2\Th)+ V_{\kk-\kkk}{\sin}^2\Th\right]\\
+ \frac S2 \sum_\kk \left(a_\mkk b_\kk+\akr_\kk \bkr_\mkk \right) \left( V_\kk-V_{\kk-\kkk} \right){\sin}^2\Th,\non 
\end{eqnarray}
$N$ is the number of spins in an AF sublattice, 
\begin{eqnarray}
g_\kk = S(J_0+J_\kk),\\
\label{d} \bD = D\left(1-1/2S \right), 
\end{eqnarray}
\begin{eqnarray}
a_\kk = \frac{1}{\sqrt{N}}\sum_n a_n \exp(i\kk \RR_{n,1}),\quad b_\kk = \frac{1}{\sqrt{N}}\sum_n b_n \exp\left[i\kk (\RR_{n,2}+{\bf Q})\right],\\
\label{jkvk} J_\kk = \sum_{\bf r} J_{\bf r} \exp(i\kk{\bf r}),\qquad V_\kk = \sum_{{\bf r}'} V_{{\bf r}'} \exp\left[i\kk({\bf r}'+{\bf Q})\right], 
\end{eqnarray}
vectors ${\bf r}$ connect sites inside one AF sublattice and vectors ${\bf r}'$ connect a site of one AF sublattice with neighboring sites of another AF sublattice. Terms ${\cal H}_3$ and ${\cal H}_4$ are also important for our consideration. We present them in \ref{appendixham} due to their cumbersomeness. One obtains the following expression for $V_\kk$ in the particular case of exchange couplings $V_u$ and $V_d$ with four upper and four lower (relative to the $z$ axis) neighboring spins: 
\begin{equation}
\label{Vk} V_\kk = 4e^{i{\bf kQ}} \cos \frac{k_x}{2} \cos \frac{k_y}{2} \left[ (V_u + V_d) \cos \frac{k_z}{2} +i (V_u -V_d)\sin \frac{k_z}{2} \right]. 
\end{equation}
In particular, it is seen from \eref{Vk} that $V_{{\bf k}_0}=0$.

It should be pointed out that the one-ion anisotropy term in \eref{hams} is a constant when $S=1/2$ so that all contributions to observable quantities from the one-ion term should vanish if $S=1/2$. In order results obtained within each order in $1/S$ to be in accordance with this requirement one has to regroup the $1/S$ series by taking into account some higher order terms in each order in $1/S$. Such a regrouping of the $1/S$ series can be done quite easily in our case. Simple analysis shows that along with each term in ${\cal H}_{0,1,2}$ proportional to $D$ there is the same term multiplied by $-1/2S$. As a result the quantity $\tilde D$ given by \eref{d}, which is zero if $S=1/2$, appears in \eref{h0}--\eref{h2} instead of $D$. \cite{chub}

Dependence of the canting angle on the field in the zeroth order of $1/S$ can be found in two equivalent ways: by minimization of the classical ground state energy ${\cal H}_0$ or by putting to zero linear terms ${\cal H}_1$. As a result we have 
\begin{eqnarray}
\label{hTh} \cos\Th &=& \left\{ 
\begin{array}{ll}
	h/h_c,& \mbox{ if } h\le h_c,\\
	1, &\mbox{ if } h>h_c, 
\end{array}
\right.\\
\label{hc} h_c &=& 2SJ_0+2S\bD+SV_0, 
\end{eqnarray}
where $h_c$ is the classical saturation field. 

It is convenient to introduce the following Green's functions: 
\begin{eqnarray}
\label{ga} G_a(k)&=&-\langle a_k \akr_k \rangle,\quad F_a(k)=-\langle \akr_{-k} \akr_k \rangle,\\
\label{gb} G_b(k)&=&-\langle b_k \bkr_k \rangle,\quad F_b(k)=-\langle \bkr_{-k} \bkr_k \rangle,\\
\label{gv} G^V(k)&=&-\langle b_k \akr_k \rangle,\quad F^V(k)=-\langle \bkr_{-k} \akr_k \rangle, 
\end{eqnarray}
where $k=(\w,\kk)$ and $a_k$ is the Fourier transform of $a_\kk (\tau)$. We have the following set of Dyson equations for these Green's functions: 
\begin{eqnarray}
\label{dyson} 
G_a(k)&=&G_0(k)\left[1 + \Sig_k G_a(k)+\Pi_k F_a(k)+\Sig^V_k G^V(k)+\Pi^V_k F^V(k)\right],\non\\
F_a(k)&=&G_0(-k)\left[\Sig_{-k}F_a(k)+\Pi_{-\bar{k}} G_a(k)+\Sig^V_{\bar{k}}F^V(k) +\Pi^V_{\bar{k}} G^V(k) \right],\\
G^V(k)&=&G_0(k)\left[\Sig^V_{-\bar k} G_a(k)+\Pi^V_{-k} F_a(k)+\Sig_k G^V(k)+\Pi_k F^V(k) \right],\non\\
F^V(k)&=&G_0(-k)\left[\Sig^V_{-k} F_a(k)+\Pi^V_{-\bar k} G_a(k)+\Sig_{-k}F^V(k)+\Pi_{-\bar k} G^V(k) \right],\non 
\end{eqnarray}
where $G_0(k)=1/(\ww)$ is the bare Green's function, $\Sig$, $\Sig^V$ and $\Pi$, $\Pi^V$ are normal and anomalous self-energy parts, respectively, and $\overline{k} = (-\w,\kk)$. One has a set of equations similar to \eref{dyson} for Green's functions $G_b(k)$, $F_b(k)$, $G^V(k)$ and $F^V(k)$ which we do not present here. Expressing $h$ via $\theta$ using \eref{hTh}, we obtain from the bilinear part of the Hamiltonian \eref{h2} for the self-energy parts in the zeroth order in $1/S$ 
\begin{eqnarray}
\label{parts} 
\Sig_{0k} &=& g_\kk\,{\cos}^2\Th + (SJ_0+S\bD)\,{\sin}^2\Th,\non\\
\Pi_{0k} &=& g_\kk\,{\sin}^2\Th - (SJ_0+S\bD)\,{\sin}^2\Th,\label{BareSelfEnergyParts}\\
\Sig^V_{0k} &=& \shalf\left[SV_\kk(1+\cos^2\Th)+SV_{\kk-\kkk}\sin^2\Th \right],\non\\
\Pi^V_{0k} &=& \shalf(SV_\kk-SV_{\kk-\kkk})\sin^2\Th.\non 
\end{eqnarray}
Notice that these bare self-energy parts do not depend on frequency. General solution of \eref{dyson} is very cumbersome. In the meantime, it can be simplified considerably when the following equalities satisfy: 
\begin{eqnarray}
\label{simple} &\Sig_{-k} = \Sig_k&,\non\\
\Pi_{-k} = &\Pi_{-\bar k} = \overline{\Pi}_k& = \overline{\Pi}_{\bar k},\\
\Sig^V_{-k} = &\Sig^V_{-\bar k} = \overline{\Sig}^V_k& = \overline{\Sig}^V_{\bar k},\non\\
\Pi^V_{-k} = &\Pi^V_{-\bar k} = \overline{\Pi}^V_k& = \overline{\Pi}^V_{\bar k}\non, 
\end{eqnarray}
which hold exactly in two important for our consideration cases. First, for bare self-energy parts \eref{parts}. Second, at $k=(0,\kkk)$ and $k=(0,0)$. It is shown below that $\kk=\kkk$ is the point at which the gap opens due to interaction between AF sublattices. As our aim is to calculate this gap in the first order in $1/S$, we assume below that \eref{simple} hold. One obtains as a result of straightforward solution of \eref{dyson} using \eref{simple} and the smallness of $V$
\begin{eqnarray}
G_a(k) &=& \left[ (\w+\Sig_k)(\w^2-\Sig^2_k+\Pi^2_k)+\Or(V^2) \right] / {\cal D} (k),\\
F_a(k) &=& \left[ -\Pi_k(\w^2-\Sig^2_k+\Pi^2_k)+\Or(V^2) \right] / {\cal D} (k),\\
G^V(k) &=& \left\{ \left[ (\w+\Sig_k)^2+\Pi^2_k \right] \bSig^V_k-2\Pi_k(\w+\Sig_k)\overline{\Pi}^V_k+\Or(V^3) \right\} / {\cal D}(k),\\
F^V(k) &=& \left[ -\overline{\Pi}^V_k(\w^2-\Sig^2_k-\Pi^2_k)-2\Sig_k\Pi_k \bSig^V_k+\Or(V^3) \right] / {\cal D}(k),
\end{eqnarray}
where 
\begin{eqnarray}
\fl {\cal D}(k) = \left(\w^2-\Sig_k^2\right)^2 + \Pi^4_k + |(\Pi^V_k)^2-(\Sig^V_k)^2|^{2}+2|\Pi^V_k|^2\left(\w^2-\Sig_k^2\right) -2|\Sig^V_k|^2\left(\w^2+\Sig_k^2\right)\non\\
+\, 2\Pi^2_k(\w^2-\Sig_k^2-|\Pi^V_k|^2-|\Sig^V_k|^2) +4\Pi_k\Sig_k(\overline{\Pi}^V_k\Sig^V_k+\Pi^V_k\bSig^V_k).\label{znamenatel} 
\end{eqnarray}
We have from \eref{parts} and \eref{znamenatel} for the bare spectrum square 
\begin{equation}
\label{barespec} 
\fl 
\epsilon^2_{\pm,\kk} = \Sig_{0\kk}^2-\Pi_{0\kk}^2 + |\Sig^V_{0\kk}|^2-|\Pi^V_{0\kk}|^2 
\pm 
\sqrt{4|\Sig_{0\kk} \Sig^V_{0\kk}-\Pi_{0\kk} \Pi^V_{0\kk}|^2-|\Sig^V_{0\kk} \Pi^{V*}_{0\kk} - \Sig^{V*}_{0\kk} \Pi^V_{0\kk}|^2}.
\end{equation}
Neglecting interaction between AF sublattices one has in \eref{barespec} $\Sig^V_{0\kk}=\Pi^V_{0\kk}=0$ and we lead to the doubly degenerate (due to two equivalent AF sublattices) spectrum of AF in magnetic field having the well-known form 
\begin{equation}
\label{specAF} \epsilon_\kk = S\sqrt{(J_0+J_\kk)\left(J_0 + J_\kk \cos2\Th + 2\bD{\sin}^2\Th \right)}. 
\end{equation}
The minimum of this spectrum is at $\kk=\kk_0$ and the spectrum is gapless at that point because $J_\kkk=-J_0$. The inter-sublattice interaction splits these two degenerate branches in the classical spectrum. In particular, it leads to the splitting of two branches at $\kk=0$ which can be observed, in particular, in ESR experiment. The value of this splitting can be found from \eref{parts} and \eref{barespec} with the result 
\begin{equation}
\label{k0} 
\delta \epsilon_{\kk=0} = \epsilon_{+,\bf 0} - \epsilon_{-,\bf 0} 
\approx 
SV_0 \frac{ 2J_0\cos^2\Th+\bD\sin^2\Th }{\sqrt{ J_0^2\cos^2\Th+J_0\bD\sin^2\Th } }.
\end{equation}

However, the inter-sublattice interaction remains the classical spectrum gapless. Really, using \eref{znamenatel} we obtain for the spectrum square at $\kk=\kkk$  
\begin{eqnarray}
\label{corrsol} 
\epsilon^2_{\pm,\kkk}
=
\left(\Sig_{\kz} \pm \Sig^V_{\kz})^2-(\Pi_{\kz} \pm \Pi^V_{\kz}\right)^2
\end{eqnarray}
and it is seen from \eref{parts} and \eref{corrsol} that $\Sig_{0\kk_0}=-\Pi_{0\kk_0}$ and $\Sig^V_{0\kk_0}=-\Pi^V_{0\kk_0}$ so that $\epsilon_{\pm,{\kk_0}}=0$. We show now that the spin-wave interaction leads to the gap in one of these branches.

\begin{figure}
\centering 
\includegraphics[scale=1.3]{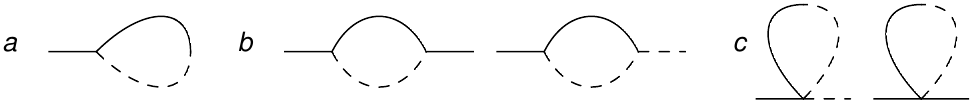} \caption{ Diagrams contributing to the gap in the first order in $1/S$. Solid and dashed lines denote Green's functions $G_a(k)$, $F_a(k)$ and $G_b(k)$, $F_b(k)$, respectively, given by \eref{ga} and \eref{gb}. Composite lines containing solid and dashed parts denote $G^V(k)$ and $F^V(k)$ given by \eref{gv}. (a) Diagrams leading to the first $1/S$ corrections to the linear term in the Hamiltonian \eref{h1}. They lead to renormalization of the angle $\Th$ that in turn leads to renormalization of the bare self-energy parts \eref{parts}. (b) Corrections to self-energy parts from three-magnon terms \eref{h3}. (c) Corrections to self-energy parts from four-magnon terms \eref{h4}. } \label{corrections} 
\end{figure}

\begin{figure}
\centering 
\includegraphics[scale=0.5]{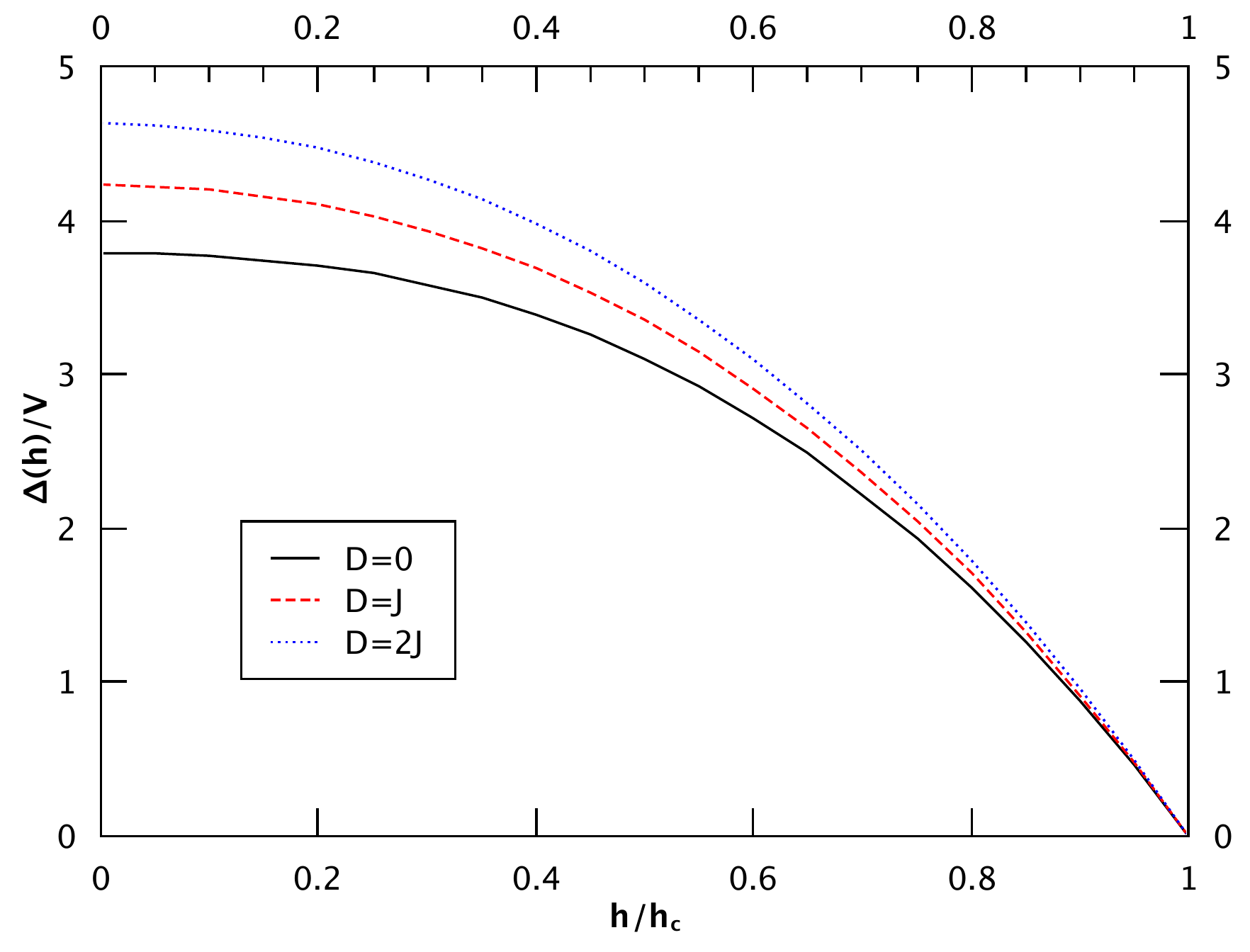} 
\caption{
\label{deltafig}
(Color online.) The field dependence of the gap $\Delta(h)$ induced by the exchange interaction $V$ between AF sublattices given by Eq.~\eref{sgap}. 
} 
\end{figure}

Diagrams leading to the gap are shown in \fref{corrections}. Corrections from them can be expressed as follows: 
\begin{equation}
\label{deltas} \Sig_\kz+ \Pi_\kz = \Sig^V_\kz+\Pi^V_\kz= S\sin^2\Th \frac{1}{N} \sum_\kk |V_\kk|^2 (\Sig_{0\kk}-\Pi_{0\kk})^2\, \epsilon^{-3}_\kk, 
\end{equation}
where $\epsilon_\kk$ is given by \eref{specAF}. It is seen from \eref{deltas} that the first nonvanishing correction at $\kk=\kkk$ is of the order of $V^2$. One concludes from \eref{parts}, \eref{corrsol} and \eref{deltas} that there is one Goldstone mode and the mode with the gap $\Delta$ for which we have 
\begin{equation}
\label{sgap} 
\Delta^2 = 4S^2\sin^4\Th(J_0+\bD) \frac1N\sum_\kk |V_\kk|^2 (\Sig_{0\kk}-\Pi_{0\kk})^2\, \epsilon^{-3}_\kk. 
\end{equation}
In particular, we recover from \eref{sgap} at $h=D=0$ the well-known Shender's expression \cite{shender}. Notice that the gap dependence on magnetic field is not trivial because the summand in \eref{sgap} is a function of $h$. We plot in \fref{deltafig} the value $\Delta(h)/V$ for some particular values of $D$ assuming that there is exchange coupling $J$ between only neighboring spins in an AF sublattice and $V_u=V_d=V\ll J$ in \eref{Vk}.

It should be stressed one more time that results obtained above are applicable for $D\lesssim J$ so that the system has long-range Neel order at $h=0$. If $D$ is sufficiently large the Neel order is destroyed by quantum fluctuations at $h<\tilde h_c<h_c$ and $h=\tilde h_c$ is a new quantum critical point. This situation is realized in DTN in which $D\gg J$. This statement can be demonstrated by the following expression for the mean value of the ion magnetization found in the first order in $1/S$ at $D\gg J$ by neglecting $V$:
\begin{equation}
\label{sz}
	\langle S^{z'} \rangle =  S - \frac{1}{\sqrt{8}N} \sum_{\bf k} \sqrt{ \frac{\widetilde{D}}{ J_0 + J_{\bf k }}}\sqrt{ 1-\frac{h}{h_c} }.
\end{equation}
For exchange interaction $J$ between only nearest neighbors we have from \eref{sz} $\langle S^{z'} \rangle \approx  S - 0.16\sqrt{(\tilde D/J)(1-h/h_c)}$. It is seen from this expression that $\tilde h_c$ is finite at $\tilde D>39JS^2$. 

We derive in the next section an expression for the gap at $h\approx h_c$ that is applicable for arbitrary $D$. It is also shown there that \eref{sgap} works qualitatively at $h\sim h_c$ even at $D\gg J$.

\section{Neighborhood of the quantum critical point $h=h_c$} 
\label{qcp}

At $h\approx h_c$ our results obtained above using the $1/S$ expansion can be supplemented by calculations using the magnon BEC theory \cite{r1,r2} which allows to obtain observable quantities as series in terms of powers of the small parameter $(h_c-h)/h_c$. As it is shown in Refs.~\cite{r1,r2}, the $1/S$ expansion works badly for $S\sim1$ at $h\approx h_c$ due to strong quantum fluctuations which are properly taken into account in the magnon BEC treatment.

Following Refs.~\cite{r1,r2} we start with the case of $h>h_c$ when all spins are parallel to the $z$ axis. Using Holstein-Primakoff representation one writes \cite{r2}
\begin{eqnarray}
\label{hpqcp}
S^{+}_{n,1} \approx \sqrt{2S} \akr_n (1- {\cal F} \akr_n a_n), \\
S^{z}_{n,1} = -S + \akr_n a_n, \\
\label{f}
{\cal F} =  1 - \sqrt{1-1/2S}
\end{eqnarray}
for the first AF sublattice and similar equalities for the second AF sublattice (with $b$ operators instead of $a$ ones). After these transformations Hamiltonian \eref{ham}--\eref{hint} has the form \eref{hamseq}, where $ {\cal H}_{1} = 0$,
\begin{eqnarray}
{\cal H}_{2} &=& \sum_{\kk} \left[ \left( \akr_{\kk} a_{\kk} + \bkr_{\kk} b_{\kk} \right) \left( g_{\kk} -\mu \right) + \left( \akr_{\kk} b_{\kk} + \bkr_{-\kk} a_{-\kk} \right) SV_{\kk} \right], \label{hamBEC:2}\\
{\cal H}_{4} &=& \sum_{1,2,3,4} \left( \akr_{1} \akr_{2} a_{3} a_{4} + \bkr_{1} \bkr_{2} b_{3} b_{4} \right) \left[ \shalf J_{4-2} - {\cal F} S \left( J_{1} + J_{4} \right) + D \right]   \non \\
&& + \sum_{1,2,3,4} \akr_{1} \bkr_{2} a_{3} b_{4}\; V_{4-2}  \non \\
&-& {\cal F} S \sum_{1,2,3,4} \left[ \left( \akr_{1} \akr_{2} a_{3} b_{4} + \akr_{4} \bkr_{3} b_{2} b_{1} \right) V_{4} + \left( \bkr_{4} \akr_{3} a_{2} a_{1} + \bkr_{1} \bkr_{2} b_{3} a_{4} \right) V_{4}^{*}\right]\label{hamBEC:4x}
\end{eqnarray}
and $\mu = h_{c} - h$ plays the role of the chemical potential \cite{r1,r2}. One obtains in this case for Green's functions $G_a(k)$ and $G^V(k)$ defined in \eref{ga} and \eref{gv}
\begin{eqnarray}
G_a(k) &=& \left( \w - \Sigma_{k} \right) / {\cal D}(k), \label{eqBEC:greensG}\\
G^{V}(k) &=& (\Sigma^{V}_{k})^{*} / {\cal D}(k),  \label{eqBEC:greensGV}\\
{\cal D}(k) &=& \left( \w - \Sigma_{k} - |\Sigma^{V}_{k}| \right) \left( \w - \Sigma_{k} + |\Sigma^{V}_{k}| \right). \label{eqBEC:Den}
\end{eqnarray}
Bare self-energy parts in these equations have the following form at $\mu<0$ (i.e., at $h>h_c$):
\begin{eqnarray}
\Sigma_{k} &=& g_{\kk} - \mu, \label{eqBEC:Self-energyS} \\
\Sigma^{V}_{k} &=& SV_{\kk}. \label{eqBEC:Self-energySV}
\end{eqnarray}
Using \eref{eqBEC:Den}, \eref{eqBEC:Self-energyS} and \eref{eqBEC:Self-energySV} we obtain the magnon spectrum containing two branches
\begin{eqnarray}
 \label{eqBEC:SimpleSpectrum}
 \epsilon_{\pm,\kk} = g_{\kk} - \mu \pm |SV_{\kk}|,
\end{eqnarray}
which is stable because $g_{\kk}\propto\kappa^2$ and $V_{\kk}\propto\kappa_x\kappa_y\kappa_z$ at $k\sim k_0$, where $\mbox{\boldmath$\kappa$}={\bf k}-{\bf k}_0$.

\begin{figure}
\centering
\includegraphics{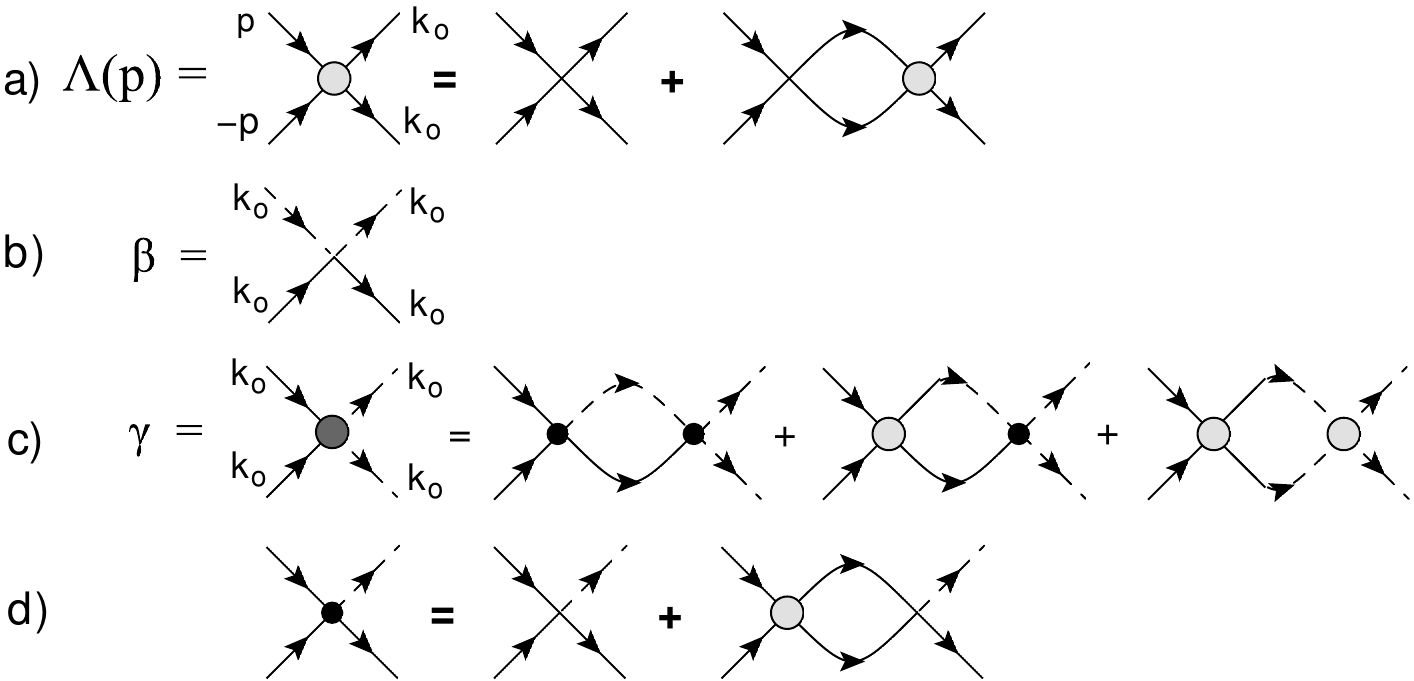}
\caption{
\label{vertexes}
Vertexes are shown schematically which appear in expression \eref{eqBEC:enery-full} for the energy. The meaning of lines (solid, dashed and composite solid-dashed) is the same as in \fref{corrections}. Plane a) represents an equation for $\Lambda(p)$ in the ladder approximation.
}
\end{figure}

As soon as $\mu$ becomes positive, the spectrum \eref{eqBEC:SimpleSpectrum} becomes unstable at ${\bf k}={\bf k}_0$ and AF long range order arises. In the magnon BEC treatment this transition is described in terms of "condensation" of magnons at states characterized by the momentum ${\bf k}_0$, i.e., $\langle a_{\kkk} \rangle$ and $\langle b_{\kkk} \rangle$ become nonzero and one has to make shifts
\begin{eqnarray}
a_{\kkk}  \rightarrow a_{\kkk} + \e^{i \phi_{a}}\sqrt{ N \p_{a} }, \quad b_{\kkk} \rightarrow b_{\kkk} + \e^{i \phi_{b}}\sqrt{ N \p_{b} }, \label{eqBEC:substitution}
\end{eqnarray}
where $\p_{a,b}$ are "condensates" densities and $\phi_{a,b}$ are phases. They should be chosen so as to minimize the energy, which has the form at $h\approx h_c$
\begin{eqnarray}
\label{eqBEC:enery-full}
E = - \mu (\p_{a}+\p_{b}) +\frac{\la}{2} (\p_{a}^2+\p_{b}^2) + \beta \p_{a} \p_{b} + 2\ga \p_{a} \p_{b} \cos[2(\phi_{a}-\phi_{b})], 
\end{eqnarray}
where $\la=2\Lambda(k_0)$, $\beta$ and $\ga$ are vertexes schematically shown in \fref{vertexes}. Vertexes $\beta$ and $\ga$ describe interaction between sublattices so that $\beta=\ga=0$ at $V=0$. Vertex $\la$ is of the zeroth order in $V$. It can be found in the ladder approximation by solving the equation schematically shown in \fref{vertexes}(a), as it was done for AF with one AF sublattice \cite{r1,r2}. In particular, $\beta$ can be approximated by its bare value, $\beta = V_{0}$, which is found from the second term in \eref{hamBEC:4x} by putting ${\bf k}_{1,2,3,4}={\bf k}_0$. Other corrections to $\beta$ can be omitted because they are of higher orders in $V$. As is seen from \eref{hamBEC:4x}, the bare value of $\gamma$ is equal to zero. As it is shown in \fref{vertexes}(c), $\gamma$ is constructed from other vertexes. We demonstrate below that $\gamma$ is the vertex that gives rise to the gap in the spectrum. It can be shown that $\ga$ is negative and of the order of $V^{2}$. Some more details on calculation of $\la$ and $\ga$ can be found in \ref{Appendix:VertexFunctions}. 

Minimizing the energy given by \eref{eqBEC:enery-full} with respect to $\p_{a,b}$ and $\phi_{a,b}$ we obtain 
\begin{eqnarray}
\phi_{a} = \phi_{b},\\
\p_{a} = \p_{b} = \p,\\
\label{mu}
\mu = (\la + \beta + 2 \gamma) \p.
\end{eqnarray}

\begin{figure}
\centering
\includegraphics{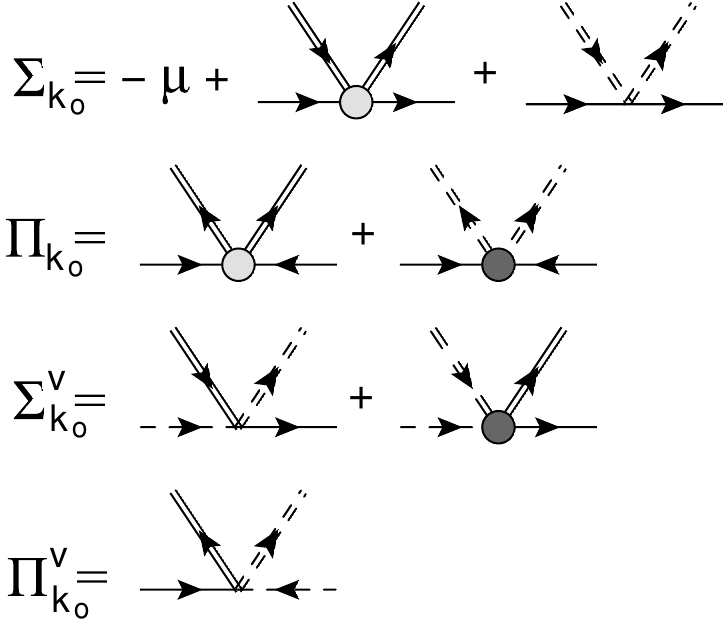}
\caption{
\label{sep}
Diagrams for self-energy parts at $\omega=0$ and ${\bf k}={\bf k}_0$ of the lowest order in $\mu/h_c$. Double straight and double dashed lines denote, respectively, condensed particles of $a$ and $b$ types and correspond to the last terms in \eref{eqBEC:substitution}. Corresponding vertexes are shown in \fref{vertexes}.
}
\end{figure}

Diagrams for self-energy parts at $\omega=0$ and ${\bf k}={\bf k}_0$ of the lowest order in $\mu/h_c$ are shown in \fref{sep}. Explicitly they can be written as follows:
\begin{eqnarray}
 \Sigma_{\kkk} = -\mu + 2 \la \p + \beta \p ,\\
 \Sigma^{V}_{\kkk} = \beta \p + 4 \ga \p ,\\
 \Pi_{\kkk} = \la \p + 2 \ga \p ,\\
 \Pi^{V}_{\kkk} = \beta \p.
\end{eqnarray}
Substituting these expressions to \eref{corrsol} and using \eref{mu} we obtain that one of the spectrum branch is gapless and we find for the gap in another branch in the leading order in $V$
\begin{eqnarray}
\label{eqBEC:Gap}
\Delta_{BEC} = 4 \mu \sqrt{ |\ga| / \la } = (h_{c}-h) \left[ 4 \sqrt{ |\ga| / \la } \right]. 
\end{eqnarray}

\begin{figure}
\centering
\includegraphics[scale=0.6]{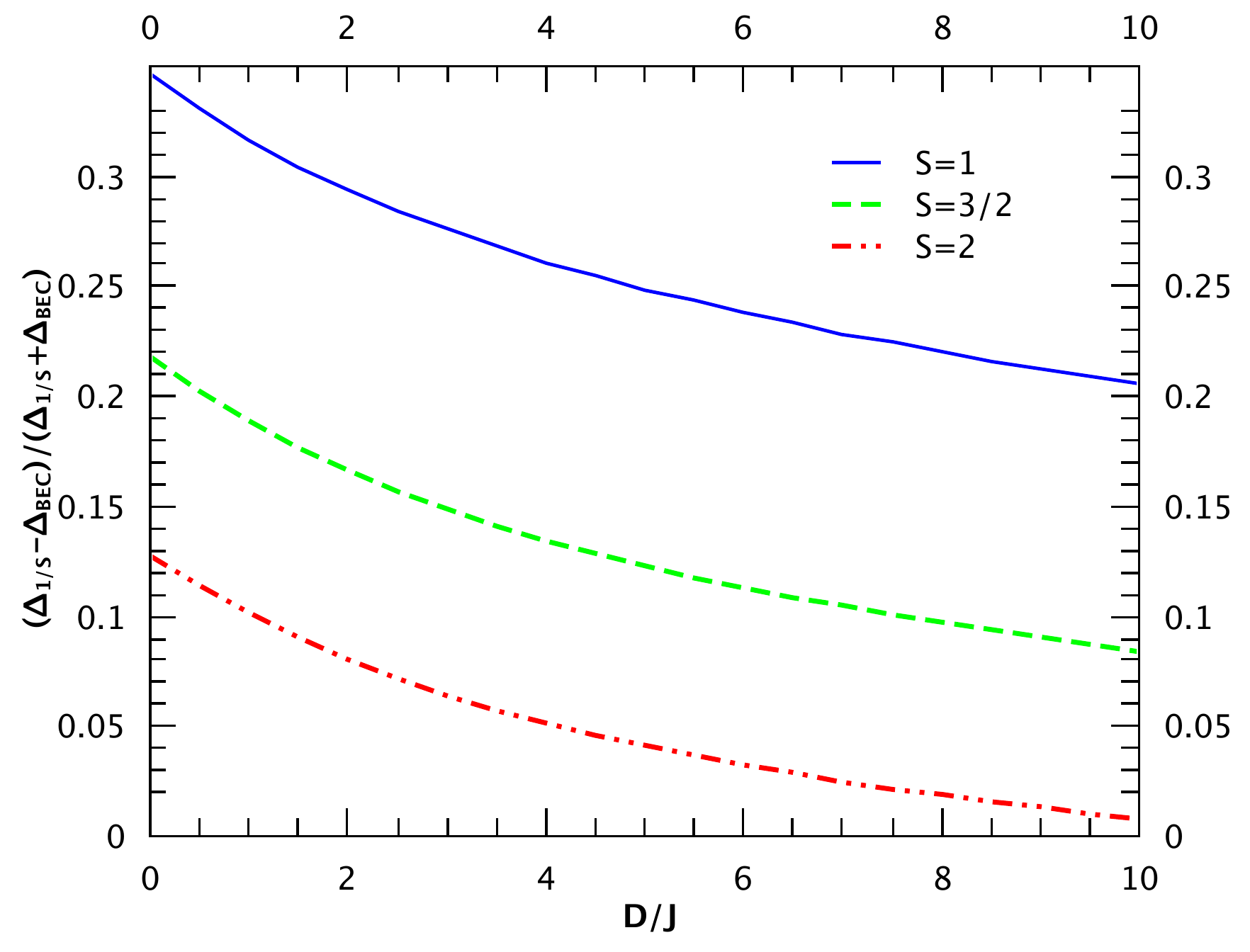}
\caption{
\label{fig:BEC-vs-SW}
Dependence of the value $(\Delta_{1/S}-\Delta_{BEC})/(\Delta_{1/S}+\Delta_{BEC})$ on $D$ and $S$ obtained using \eref{eqBEC:Gap} and \eref{eqBEC:GapSW}.
}
\end{figure}

As is pointed out above, results obtained in the magnon BEC treatment are more reliable than those of the $1/S$ expansion at $h\approx h_c$ if $S\sim1$, because fluctuations are properly taken into account in the BEC theory. It is interesting to compare results obtained within these two approaches at $h\approx h_{c}$. Expression \eref{sgap} for the gap reproduces qualitatively the field dependence of the gap because it also has a linear dependence on the field at $h\approx h_{c}$:
\begin{eqnarray}
\label{eqBEC:GapSW}
\Delta_{1/S} \approx (h_{c}-h) \left[  \frac{4S}{h_{c}} \sqrt{ (J_{0}+\bD) \frac1N\sum_\kk |V_\kk|^2 (\Sig_{0\kk}-\Pi_{0\kk})^2\, \epsilon^{-3}_\kk} \right]_{h=h_{c}}.
\end{eqnarray}
But the coefficients in \eref{eqBEC:Gap} and \eref{eqBEC:GapSW} are different. We plot the value $(\Delta_{1/S}-\Delta_{BEC})/(\Delta_{1/S}+\Delta_{BEC})$ in \fref{fig:BEC-vs-SW} for three spin values, $S=1$, 3/2 and 2. It is seen from \fref{fig:BEC-vs-SW} that results obtained within these two approaches become closer as $S$ rises as it must be. Notice also that the $1/S$ expansion overestimates the gap value. In particular, at $S=1$ and $D=0$ we have $\Delta_{1/S}\approx2\Delta_{BEC}$.  Particular calculations show also that the difference between $\Delta_{1/S}$ and $\Delta_{BEC}$ is larger in quasi-low-dimensional systems.

\section{Discussion} 
\label{dtn}

To the best of our knowledge the only compound described by the Hamiltonian \eref{ham}--\eref{hint} which has been studied experimentally in a magnetic field is DTN. Unfortunately our results obtained using the $1/S$ expansion are not entirely applicable to DTN because the one-ion anisotropy is very large in this material. As a result the ground state at small field is that with $S^z=0$ and the Neel ordering arises in the interval $\tilde h_c<h<h_c$, where $\tilde h_c\approx2.1$ T and $h_c\approx12.6$ T is the saturation field. Based on the above consideration of the neighborhood of the point $h=h_c$ one could expect that the results obtained using the $1/S$ expansion are valid qualitatively near and below $h_c$. Unfortunately the data of ESR experiment in DTN \cite{1} are available only for $h<10\,{\rm T}\approx0.8h_c$ which makes impossible a quantitative description using the expressions obtained above. Thus, we can make only a number of qualitative statements. According to the crystal structure analysis \cite{cryst} one could expect $V_u=V_d=V$ in DTN. Equations \eref{Vk} and \eref{k0} with $V\sim 0.1$ K give the optical mode splitting of the same order as that observed in ESR experiment \cite{1} at $h\approx10$ T. Estimations made using \eref{sgap} show that $V\sim 0.1$ is required also to describe the gap at $h\approx10$ T observed experimentally. 

Expression \eref{eqBEC:Gap} for the gap at $h\approx h_c$ can be useful in further experimental discussion of DTN. Values of $\la$ and $\ga$ obtained for DTN as it is described in \ref{Appendix:VertexFunctions} are the following (general expressions for them are very cumbersome):
\begin{eqnarray}
\label{dtnvert}
\la_{DTN} &=& 4.6 \, {\rm K},\nonumber\\
\frac{\ga_{DTN}}{V^2} &=& - 0.14\, {\rm K^{-1}}.
\end{eqnarray}

\section{Conclusion} 
\label{conc} 

To conclude, we discuss using the $1/S$ expansion a Heisenberg antiferromagnet containing two interpenetrating AF sublattices (see \fref{affig}) coupled by a small exchange interaction $V$ which Hamiltonian is given by \eref{ham}--\eref{hint}. The classical spectrum contains a doubly degenerate Goldstone mode if one neglects the interaction between AF sublattices. In zeroth order in $1/S$ the exchange interaction between sublattices gives rise to the optical mode splitting at ${\bf k}=0$, which can be measured in ESR experiment and which value is given by \eref{k0}, whereas the classical spectrum remains gapless at ${\bf k}={\bf k}_0$. In accordance with the previous result by Shender \cite{shender} we observe that quantum fluctuations lead to the gap in one of these two modes whose value is given by \eref{sgap} in the first order in $1/S$. \Eref{sgap} is a generalization of the Shender's expression for a finite magnetic field.

We calculate also the gap value at $h\approx h_c$ in the leading order of small parameter $(h_c-h)/h_c$ using the magnon BEC theory (see \eref{eqBEC:Gap}). By comparing this result with that obtained within the $1/S$ expansion we conclude that the latter approach gives qualitatively correct result at $h\sim h_c$ even at large one-ion anisotropy $D$ but it overestimates the gap value (see \fref{fig:BEC-vs-SW}).

Particular estimations using \eref{k0} and \eref{sgap} show that $V\sim0.1$ K is needed to describe the optical mode splitting and the gap obtained in DTN in recent ESR experiment \cite{1}. Expression \eref{eqBEC:Gap} for the gap at $h\approx h_c$ with values of $\la$ and $\ga$ given by \eref{dtnvert} can be useful in further experimental discussion of DTN.

\ack

We are thankful to Prof.~A.~I.~Smirnov for useful discussion. This work was supported by RF President (grant MK-329.2010.2), RFBR grant 09-02-00229, and Programs "Quantum Macrophysics", "Strongly correlated electrons in semiconductors, metals, superconductors and magnetic materials" and "Neutron Research of Solids".

\appendix

\section{Expressions for ${\cal H}_3$ and ${\cal H}_4$} 
\label{appendixham}

We have for ${\cal H}_3$ and ${\cal H}_4$ in \eref{hamseq} 
\begin{eqnarray}
\fl \sqrt{N}{\cal H}_3 = i\frac{\sqrt{S}}{4\sqrt{2}}\sin\Th\cos\Th \sum_{1,2,3} (\akr_1 \akr_2 a_3 -\akr_3 a_2 a_1)\left(2J_0-8J_1+10D-\frac{h}{S\cos\Th}+V_0\right) \non\\
\label{h3} {}- i\sqrt{\frac{S}{2}}\sin\Th\cos\Th \sum_{1,2,3} (\akr_1 \bkr_2 b_3+\akr_3 a_2 b_1)(V_1-V_{1-\kkk})+h.c.,\\
\fl N{\cal H}_4 = \sum_{1,2,3,4} \akr_1 \akr_2 a_3 a_4 \left(\frac{1-2{\sin}^2\Th}{2}J_{4-1}-\frac{{\cos}^2\Th}{4}(J_1+J_4)+D(1-\frac{3}{2}{\sin}^2\Th)\right)\non\\
+\sum_{1,2,3,4} (\akr_1 \akr_2 \akr_3 a_4+\akr_4 a_3 a_2 a_1)\left(-\frac{{\sin}^2\Th}{4}J_1+\frac{D{\sin}^2\Th}{4}\right)\non\\
+\sum_{1,2,3,4} \akr_1\bkr_2 a_3 b_4 \left({\cos}^2\Th V_{4-2}+{\sin}^2\Th V_{4-2+\kkk}\right)\non\\
-\frac{1}{8}\sum_{1,2,3,4} (\akr_1\akr_2 a_3 b_4+\akr_4\bkr_3 b_2 b_1)\left[(1+{\cos}^2\Th)V_4+{\sin}^2\Th V_{4-\kkk}\right]+h.c. \non\\
\label{h4} -\frac{1}{8}\sum_{1,2,3,4} (\akr_1\bkr_2\bkr_3 b_4+\akr_4 a_3 a_2 b_1)\left( V_1-V_{1-\kkk} \right) {\sin}^2\Th+h.c., 
\end{eqnarray}
where momentum conservation laws $\sum_{i=1}^3{\bf k}_i=\kkk$ and $\sum_{i=1}^4{\bf k}_i=0$ are implied in sums of \eref{h3} and \eref{h4}, respectively. 

\section{Vertices at $h\approx h_c$} 
\label{Appendix:VertexFunctions}
 
We derive in this appendix vertexes $\Lambda(\omega=0,{\bf p})$ and $\ga$ diagrams for which are shown in \fref{vertexes}(a) and (c), respectively. Let us start with $\Lambda(\omega=0,{\bf p})$. The main contribution to it is given by ladder diagrams which sum can be found by solving the integral equation shown schematically in \fref{vertexes}(a). This equation can be solved easily using the substitution
\begin{eqnarray}
\label{eqBECApp:VertexThroughAB}
\Lambda(\omega=0,{\bf p}) = A + B J_{\pp},
\end{eqnarray}
where $A$ and $B$ are constants. Substituting \eref{eqBECApp:VertexThroughAB} into the equation we obtain the following set of algebraic equations on $A$ and $B$:
\begin{eqnarray}
\label{sys}
 A \left( x + \frac{ 1 - {\cal F} S }{ D + {\cal F} S J_{0} } \right) + B y = 1,\\
 \frac{1}{J_{0}} A \left( y ( 1 - {\cal F} S ) - {\cal F} S \right) + B \left( 1 - y ( 1 - {\cal F} S) \right) 
 = - \left( \frac12 + {\cal F} S \right),\non
\end{eqnarray}
where ${\cal F}$ is given by \eref{f} and
\begin{eqnarray}
x = \NN \sum_{\kk} \frac{1}{ SJ_{0} + SJ_{\kk} },\\
y = \NN \sum_{\kk} \frac{ J_{\kk} }{ SJ_{0} + SJ_{\kk} } = \frac{1}{S} - J_{0} x. \non
\end{eqnarray}

The general solution of the system \eref{sys} is rather cumbersome. But it can be brought into the following compact form in the particular case of $S=1$:
\begin{eqnarray}
\label{ab}
A &=& J_{0} \frac{ ( 1 + J_{0} x ) \left[ D + J_{0} (1-\frac{1}{\sqrt{2}})\right] }{J_{0} x ( 3 J_{0} + 4D) - 2D - J_{0}  }, \\
B &=& \frac{ -J_{0} (1+\frac{1}{\sqrt{2}}) - 2D + J_{0} x \left[J_{0} (1-\frac{1}{\sqrt{2}}) +D\right] }{ J_{0} x ( 3 J_{0} + 4D) - 2D - J_{0} }. \non
\end{eqnarray}
As a result we have at $S=1$ from \eref{eqBECApp:VertexThroughAB} and \eref{ab}
\begin{eqnarray}
\la = 2\Lambda(\omega=0,\kkk) = 2(A-BJ_{0}) = J_{0} \frac{ 2J_{0} + 3D }{ J_{0} x ( 3 J_{0} + 4D) - 2D - J_{0}}.
\end{eqnarray}

To calculate $\ga$ we have to take into account diagrams shown in \fref{vertexes}(c) which involve the vertex $\Lambda(\omega=0,{\bf p})$ obtained above and the vertex denoted by the solid circle. Diagrams for the latter vertex are presented in \fref{vertexes}(d). The result of their summation can be expressed as a multiplication of the bare value of that vertex given by the last term in \eref{hamBEC:4x} by the constant 
\begin{eqnarray}
c = 1 - \frac1N \sum_{\kk} \frac{ \Ga_{\kk}}{g_{\kk}}.
\end{eqnarray}
As a result we obtain
\begin{eqnarray}
\ga = - \frac1N \sum_{\kk} \frac{ |V_{\kk}|^2 }{ 32 g_{\kk}^3 } \left( {\cal F} S c g_{\kk} + 4 S \La(\kk) \right)^{2}.
\end{eqnarray}

If there are different exchange constants $J_x$, $J_y$ and $J_z$ along different directions (as in DTN, where $J_x=J_y\ne J_z$) one has to trial the solution of the equation for $\Lambda(\omega=0,{\bf p})$ in the form
\begin{eqnarray}
\Lambda(\omega=0,{\bf p}) = A + B_x J_{xp_x} + B_y J_{yp_y} + B_z J_{zp_z}.
\end{eqnarray}
The resultant expressions are very cumbersome in this case and we do not present them here. They lead to values for $\lambda$ and $\ga$ given by \eref{dtnvert} with DTN parameters $S=1$, $D=8.9$~K, $J_z=2.2$~K and $J_{x,y}=0.18$~K \cite{6}.

\section*{References}

\bibliography{TS} 

\end{document}